\documentclass[10pt,journal]{IEEEtran}

\usepackage{cite}
\usepackage{amsmath,amsfonts}
\usepackage{algorithmic}
\usepackage{array}
\usepackage{url}
\usepackage[hidelinks]{hyperref}
\usepackage{color}
\usepackage{comment}
\usepackage{graphicx}
\begin{document}

\title{Delay Analysis of EIP-4844}

\author{Pourya~Soltani and Farid~Ashtiani
        \IEEEcompsocitemizethanks{\IEEEcompsocthanksitem The authors are with the Department of Electrical Engineering, Sharif University of Technology (SUT). Tehran 11155-4363, Iran (Email: pourya.soltani@sharif.edu, ashtianimt@sharif.edu)}
}

\IEEEtitleabstractindextext{%
\begin{abstract}
Proto-Danksharding, proposed in Ethereum Improvement Proposal 4844 (EIP-4844), aims to incrementally improve the scalability of the Ethereum blockchain by introducing a new type of transaction known as blob-carrying transactions. 
These transactions incorporate binary large objects (blobs) of data that are stored off-chain but referenced and verified on-chain to ensure data availability. 
By decoupling data availability from transaction execution, Proto-Danksharding alleviates network congestion and reduces gas fees, laying the groundwork for future, more advanced sharding solutions. 
This letter provides an analytical model to derive the delay for these new transactions. 
We model the system as an $\mathrm{M/D}^B/1$ queue which we then find its steady state distribution through embedding a Markov chain and use of supplementary variable method.
We show that transactions with more blobs but less frequent impose higher delays on the system compared to lower blobs but more frequent. 
\end{abstract}

}

\maketitle

\IEEEdisplaynontitleabstractindextext

\IEEEpeerreviewmaketitle

\section{Introduction}

\IEEEPARstart{B}{lockchain} technology, while revolutionary, faces significant challenges in terms of scalability. 
As the popularity of blockchain-based applications has grown, so too has the demand for network resources. 
This has resulted in increased congestion and higher transaction fees on many popular networks, particularly Ethereum. 
The scalability issue fundamentally arises from the need to maintain the decentralized nature of these networks while trying to process an ever-growing number of transactions. 
While the long-term solution lies in data sharding \cite{Danksharding}, Ethereum has adopted a rollup-centric roadmap as the midterm scaling solution to scalability challenges \cite{rollup-roadmap}.  
This strategic approach places rollups at the forefront of Ethereum’s scaling solutions.

Rollups work by processing transactions off-chain and then posting summarized batches of these transactions back to the main chain for final verification. 
This approach significantly reduces the amount of data that needs to be processed and stored on the main chain, thereby alleviating congestion and lowering transaction fees. 
By moving the bulk of transaction processing off-chain, rollups can greatly enhance the throughput of blockchain networks without sacrificing security. 

Ethereum Improvement Proposal (EIP) 4844, also known as Proto-Danksharding, introduces a new transaction type specifically designed to enhance the efficiency and scalability of rollups. 
These transactions, also called blob-carrying transactions (BTXs), enable the inclusion of binary large objects (blobs) of data for rollups to publish their data.
This proposal focuses on reducing the cost and complexity associated with rollup transactions by enabling more efficient data availability solutions. 
EIP-4844 aims to bridge the gap between current rollup technologies and future sharding implementations, providing an intermediate scaling solution that can be deployed in the near term. 
In particular, EIP-4844 implements the same transaction format as future sharding and establishes a gas fee market for this type of transaction. 
However, unlike full data sharding, EIP-4844 does not shard transactions \cite{EIP-4844}.

Given that EIP-4844 is still in its early stages, research on its implications remains limited.
Almost a year prior to its implementation, projections in \cite{crapisProjection} estimated that the demand for blobs then was approximately ten times lower than the sustainable target, with an anticipated time frame of one to two years required to reach this level.
Until the demand aligns with the sustainable target, the data gas price is expected to remain near zero.
The authors in \cite{crapis2023eip} propose an economic model to analyze implications of EIP-4844 for rollup data posting strategies. 
In their model, rollup costs consist of data posting costs and delay costs. 
They identify a threshold for the transaction arrival rate below which rollups prefer the existing gas market, while higher rates make the new data market more appealing. 
Additionally, the study finds that shared blob posting can reduce costs, but its benefits depend on the transaction volumes of the participating rollups. 

In an interesting work in \cite{park2024impact}, impact of EIP-4844 on different aspects of Ethereum blockchain has empirically been analyzed.
The authors conducted extensive data collection.
It has been shown that the blobs make the synchronization slower leading to higher fork rates, highlighting the negative impact of the proposal on consensus security.
Nevertheless, their findings confirm that the reduced fees resulting from EIP-4844 has in fact incentivized a greater volume of transactions on rollups.

In this letter, we aim to derive the average delay of BTXs in the blockchain.
BTXs are issued from different rollups which are then served in bulk of size at most $B$ at each block instance.
Our contributions are as follows.
\begin{itemize}
	\item We model the process of BTX service as an $\mathrm{M/D}^B/1$ queue which we then solve through embedding a Markov chain at departure instants and deriving the steady state probability distribution through introducing elapsed time since service as a supplementary variable. 
	\item We show that less frequent larger BTXs impose higher delays on the system compared to more frequent but smaller ones.
		So to estimate when we reach the sustainable target, one needs to know how blob usage in BTXs evolves as well.
	\item Analyzing data from three months of Ethereum blocks, we find that single blob BTXs significantly outnumber larger BTXs.
\end{itemize}

The organization of this letter is as follows. 
In Section \ref{Background} we review some relevant background on Ethereum.
In Section \ref{Model} we introduce our model.
Then in Section \ref{Results} we report some numerical results.
We conclude the letter in Section \ref{Conclusion}.

\section{Background}\label{Background}

\subsection{Ethereum Scalability Issues}

The scalability trilemma in blockchain technology states that it is difficult to simultaneously achieve decentralization, scalability, and security. 
To address the scalability issue, various solutions, categorized as Layer 1 and Layer 2, have been proposed \cite{zhou2020solutions}. 
Ethereum's long-term roadmap prioritizes data sharding to enhance the network capacity to store and manage data efficiently while leveraging rollups to handle transaction processing. 
While data sharding organizes transaction data storage, the actual execution of transactions will involve rollups and other Layer 2 solutions \cite{Danksharding}. 
Proto-Danksharding (EIP-4844) represents an incremental step towards full data sharding, offering immediate scalability improvements by providing more space for blobs of data rather than traditional transactions \cite{EIP-4844}. 
This further aligns with Ethereum’s current rollup-centric roadmap, which focuses on enhancing data availability for rollups without the Ethereum protocol interpreting the data directly \cite{rollup-roadmap}.

\subsection{Rollups}

Rollups are a Layer 2 scaling solution for Ethereum that enhance transaction throughput and reduce costs through a two-layer model involving off-chain execution and on-chain settlement.
In this system, transactions are processed off-chain and then aggregated into batches, which are posted to a smart contract on the Ethereum mainnet \cite{thibault2022blockchain}. 
This significantly reduces transaction fees and alleviates the load on the base layer, thereby enhancing scalability. 
Rollups inherit Ethereum’s security properties by recording their results on the mainnet, ensuring that off-chain computations remain secure and verifiable.

Rollups come in two main types: optimistic rollups and zk-rollups \cite{thibault2022blockchain}. 
Optimistic rollups assume transactions are valid by default and only perform computations in the event of a dispute, relying on a challenge period during which fraudulent transactions can be contested. 
Zk-rollups, on the other hand, use zero-knowledge proofs to verify the correctness of transactions, providing immediate finality and higher security assurances without the need for a challenge period. 
Both types of rollups contribute to scaling Ethereum by moving computation off-chain while maintaining on-chain security.


\subsection{Proto-Danksharding (EIP-4844)}

Proto-Danksharding, a step toward full data sharding, introduces the concept of data blobs, i.e., binary large objects that are referenced in transactions but not executed on-chain \cite{EIP-4844}. 
These blobs are used primarily by Layer 2 rollup protocols to support high-throughput transactions. 
This approach aims to reduce gas fees and alleviate network congestion by improving data handling efficiency.
The key innovation is that while the blobs themselves are not executed on-chain, their availability must be verified. 
Validators and users are required to ensure that these blobs can be downloaded and accessed from the network, thereby confirming their availability.

EIP-4844 proposes a new transaction type known as blob-carrying transactions, which include these data blobs. 
Validators are tasked with storing and propagating these blobs, ensuring their availability without processing the data on-chain. 
This separation of data availability from execution enhances scalability. 
To safeguard the integrity and availability of the data blobs, the proposal employs Kate-Zaverucha-Goldberg (KZG) commitments.
The blobs are then represented by a versioned hash of the blob’s KZG commitment hash \cite{EIP-4844, EIP-4844-FAQ}. 
While blobs are deleted from the consensus layer after a specified amount of time, their versioned hashes remain to ensure data integrity and availability verification.

As a consequence of the new transaction type, a multidimensional fee market is implemented \cite{EIP-4844-FAQ}.
This market categorizes transactions and data blobs into various dimensions, each with its own pricing mechanism, reducing congestion and promoting a more balanced fee structure. 
By differentiating fees based on data blob types and sizes, Proto-Danksharding lowers transaction costs and enhances scalability.



\section{System Model \& Analysis}\label{Model}

BTXs usually arrive from multiple independent sources with different demand and hence posting strategies \cite{l2beat}.  
Accordingly, we consider arrival of BTXs to follow a Poisson point process with rate $\lambda$. 
We consider multi-dimensional fees are such that all BTXs are eligible for service.
Nevertheless, we overlook the priority in BTXs since we are only after the average delay.
We further consider only the single blob BTXs in our analysis though we discuss the impact of higher blob usage in Section \ref{Results}.
Service is deterministic with duration $\tau$ as the consequence of the Merge\footnote{The Merge refers to Ethereum's transition from a proof-of-work (PoW) consensus mechanism to a proof-of-stake (PoS) system. Blocks are produced roughly every 12 seconds in PoS.} and $B$ is the maximum number of blobs per block.
The above description then resembles the characteristics of an $\mathrm{M/D}^B/1$ queue as a proper model for a shard. 
Nonetheless, there are subtleties at play.
First, the service is pegged to the main chain's consensus, i.e., the chain grows irrespective of request for blobs.
In other words, service points occur regularly whether there are any BTXs or not.
Second, we are dealing with what has been referred to as a modified batch service in \cite{bhat1964imbedded}, which means inputs can join an incomplete batch during the service epoch.

To obtain the system characteristics of our model, we need to derive the steady state solution of the system.
Throughout this letter we consider stability for our $\mathrm{M/D}^B/1$ queue, i.e., $\rho = \lambda \tau / B < 1$.
We could then describe the state of our system as $(n, x)$ with $n$ as the number of BTXs in the queue (both waiting and in service) and $x$ the elapsed time since the last service.  
Nonetheless, to simplify the description and analysis we begin by embedding a Markov chain at departure instants. 
Accordingly, the new states will become the number of remaining BTXs seen in the queue upon block production.

Introducing $\alpha_k = \frac{e^{-\lambda \tau} (\lambda \tau)^k}{k!}$ as the probability of $k$ arrivals during a service epoch and $\beta_j = \sum_{k=0}^{j}{\alpha_k}$,
\begin{equation}\label{TPM}
	P = 
	\begin{bmatrix}
		\beta_{B} & \alpha_{B+1} & \alpha_{B + 2} & \alpha_{B+3} & \alpha_{B+4} & \ldots \\
		\beta_{B-1} & \alpha_{B} & \alpha_{B + 1} & \alpha_{B+2} & \alpha_{B+3} & \ldots \\
		\vdots & \vdots & \vdots & \vdots & \vdots & \vdots \\
		\beta_{1} & \alpha_{2} & \alpha_{3} & \alpha_{4} & \alpha_{5} & \ldots \\
		\alpha_0 & \alpha_{1} & \alpha_{2} & \alpha_{3} & \alpha_{4} & \ldots \\
		0 & \alpha_{0} & \alpha_{1} & \alpha_{2} & \alpha_{3} & \ldots \\
		0 & 0 & \alpha_{0} & \alpha_{1} & \alpha_{2} & \ldots \\
		\vdots & \vdots & \vdots & \vdots & \vdots & \vdots
	\end{bmatrix},
\end{equation}
would be the transition probability matrix of the embedded Markov chain.
The elements of $P$, $P_{ij}$s for $i, j >=0$, then show the transition probability from state $i$ to state $j$ after a departure instant.
Two states are not reachable, i.e., $P_{ij}=0$, when $j < i - B$ since at most $B$ BTXs can be served at each epoch.
For the case of $j>0$, $B$ BTXs would have left the system and hence $P_{ij} = \alpha_{j - i + B}$ when $i, j$ are reachable.
For the case of $j=0$, when $i, j$ are reachable, i.e., $i <= B$, any less than $B - i$ arrivals will lead to an empty queue at the next departure instant which then lead to $P_{ij} = \beta_{B-i}$.

Solving for $\pi^+ P = \pi^+$ where $\pi^+ = [\pi^+(0), \pi^+(1), ...]$ along with $\sum_{n=0}\pi^+(n) = 1$ then gives $\pi^+(n)$ as the probability that the departure sees $n = 0, 1, ...$ BTXs in the queue.
It is worth mentioning that we could also incorporate monetary policies like (exponential) EIP-1559 into our model.
EIP-1559 features a base fee that adjusts dynamically with network congestion.
Upon departure, base fee changes which affects not only the demand (input rate) but also the queue length (state).
Nonetheless, this is left for the future.

Due to the batch departure, the states that arrivals observe will not be the same as departures.
We can nevertheless obtain the time-stationary distribution of the system by deriving them for random instances in time. 
To do so, we need to use supplementary variables to keep track of additional information about the system state \cite{cox1955analysis}.
In particular, we just need to know how the departure point distribution can change in an elapsed time $x$.
We then have
\begin{equation}\label{ranDist}
	\overline{\pi}(n) = \sum_{k=0}^n \frac{\pi^+(k)}{\tau} \int_0^\tau \frac{e^{-\lambda x} (\lambda x)^{n-k}}{(n-k)!} \, dx,
\end{equation}
where $\overline{\pi}(n)$ would be the probability of $n = 0, 1, ...$ BTXs in the system at any random instance in time.
As \eqref{ranDist} suggests, random time states are reached from lower departure time states through the expected arrivals in the elapsed time.

Since the above random points are equivalent to a Poisson point process \cite{kleinrock1974queueing} and Poisson arrival sees time averages (PASTA), the steady state distribution $\pi(n)$ will be equal to the time-stationary distribution $\overline{\pi}(n)$.
However, in order to be able to solve the balance equations, we cannot have the states till infinity. 
Hence, we place a limit $n_{max}$ for the number of states in \eqref{TPM}.
Accordingly, a normalization factor will then be required in \eqref{ranDist}.
Nevertheless, in case the limit is large enough, the normalization factor approaches one and \eqref{ranDist} will be a good approximation for steady states.

Finally, with $N = \sum_{n=0}^{n_{max}}n\pi(n)$ as the average number of BTXs in the system, through Little's law we obtain $T = N/\lambda$ as the average time spent by BTXs in the system.

\section{Numerical Results}\label{Results}

Since we are in the initial stages of EIP-4844 deployment, the usage of blobs is relatively low. 
As a result, the cost of the blob itself is very low and is estimated to remain as such for about one to two years \cite{crapisProjection}. 
We have further collected data for blob usage from June 1st to August 31st 2024, covering Ethereum blocks from \#19993250 to \#20651993 \cite{github}.
Approximately 34\% of blocks in that range contain no BTXs.
The average BTX rate is 1.33 per block with blob usage as Table \ref{table:blob_lengths_rates}.
Single blob BTXs dominate the market.
As derived from Table \ref{table:blob_lengths_rates}, the average blob usage per BTX is 1.88 resulting in an average blob rate of 2.51 per block.
The current target and maximum number of blobs per blocks are three and six, respectively \cite{EIP-4844}.
This implies that there are no noticeable market-induced delays or queueing delays, only service time.
In other words, the demand is not sufficient for a queue to be formed.
Hence, due to lack of real world data, we compare our results to an event-based simulation to validate them.

\begin{table}[t!]
	\centering
	\caption{Percentage of BTXs with different blob usage}
	\label{table:blob_lengths_rates}
	\begin{tabular}{|c|c|c|c|c|c|c|}
		\hline
		\textbf{Number of blobs} & 1    & 2    & 3    & 4    & 5    & 6    \\ \hline
		\textbf{BTX share}       & 71.73 & 3.16 & 9.49 & 0.14 & 11.71 & 3.77 \\ \hline
	\end{tabular}	
\end{table}

As mentioned, currently the target and maximum number of blobs per blocks are three and six, respectively.
However, there were discussions and initial proposals suggesting other values for these parameters. 
Specifically, early considerations included targets and maximums of respectively two and three \cite{crapisProjection} or eight and sixteen \cite{EIP-4844-OLD} blobs per block.
Hence, we include them to see their effect on delay.
We set our block production rate the same as Ethereum's, i.e., $\mu = 1 / \tau = 1 / 12$.
We would have then swept $\lambda$ to obtain delay, but since different $B$s lead to different stability regions, for the sake of comparison we'd rather sweep $\rho = \frac{\lambda}{B \mu}$.

\begin{figure}[t!]
	\centering
	\includegraphics[width=1.0\linewidth]{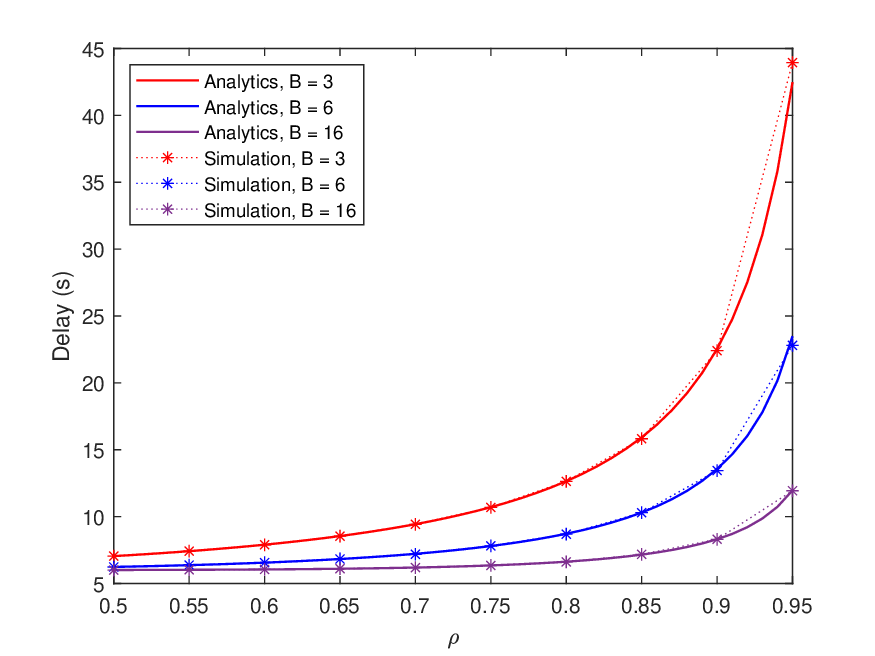}
	\vspace{-0.65cm}
	\caption{Average delay of BTXs in a shard for different loads.}
	\label{fig:delayVSrho}
	\vspace{-0.3cm}
\end{figure}

Fig. \ref{fig:delayVSrho} shows the delay in a shard for different loads along with simulation results at 0.05 steps in $\rho$. 
As load in the system decreases, delay approaches service time which is $S = \frac{m_2}{2m_1} = \tau/2$ where $m_k$ is the kth moment of inter block times \cite{kleinrock1974queueing}.
When the load in the system increases, delay increases as well.
However, exponential EIP-1559 will try to keep $N$ equal to the target by increasing the base fee and consequently reducing demand.
Nonetheless, when the demand rises beyond the target, whether due to prices changing channel or naturally through time, blobs can even experience some delays multiple of block intervals.

For the case of $B = 6$ we have $N = 3$ when $\rho \approx 0.76$. 
Comparing to the current blob demand, we'd have $\rho \approx 0.42$, if all the blobs were issued singly.
So, almost twice the current demand would suffice to reach the target. 
But, as Table \ref{table:blob_lengths_rates} shows, not all the BTXs are single blobs. 
Some rollups prefer to submit BTXs with larger number of blobs.
This is usually to gain better compression efficiency \cite{BatchCompression}.
To account for the general case, we need to introduce new classes of BTXs based on their blob usage.
This would then add more dimensions to our Markov chain further complicating the analysis.
Instead, we'd rather discuss the impact of such classes through $B$.

\begin{figure}[t!]
	\centering
	\includegraphics[width=1.0\linewidth]{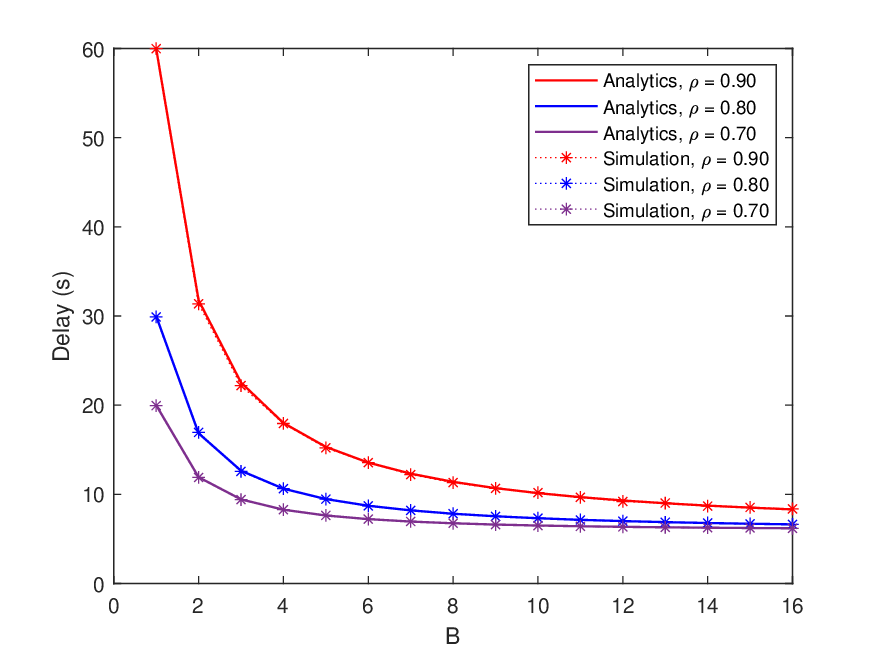}
	\vspace{-0.65cm}
	\caption{Average delay of BTXs in a shard with different blob capacities.}
	\label{fig:delayVSbatch}
	\vspace{-0.2cm}
\end{figure}

In this respect, BTXs with more blobs mean less than $B$ BTXs can be served in every block.
In our model this means that the effective batch size, $B_{eff}$, is less than the maximum allowable blobs per block, $B$. 
Fig. \ref{fig:delayVSbatch} further highlights the effect of $B_{eff}$ on delay.
As it can be seen, higher $B$ incurs less delay at the same load. 
Consider for the same $\rho = \frac{\lambda}{B \mu}$, instead of single blob BTXs with rate $\lambda$, incoming BTXs used up all the blobs with rate $\lambda/B$.
It is like they wait till $B$ blobs are present then submit them as a single BTX.
It can be seen that in this case delay increases dramatically.

As observed in Fig. \ref{fig:delayVSbatch}, delay is a piecewise linear convex function in $B$ and by Little's law, so will be the average number of blobs in the system. 
So to estimate when we reach the target, one needs to know how blob usage in BTXs evolves as well.
Consequently, as rollups strive for larger BTXs, we would require much less than twice the current demand to reach the target and hence a market is formed.

It is worth mentioning that the introduction of shared sequencing would probably have the same impact as using up all the blobs available though the input distribution might not necessarily follow Poisson.
Shared sequencing aims to decentralize the transaction ordering process, mitigating the risk of centralization in rollup operation and enhancing the robustness of the entire network.
Accordingly, a shared sequencer in charge of multiple rollups will produce BTXs comprised of more blobs \cite{Ponyo}, causing increase in overall delay. 

\section{Conclusion \& Future Works}\label{Conclusion}

EIP-4844 represents a significant advancement in addressing Ethereum's scalability challenges by introducing blob carrying transactions.
We proposed an $\mathrm{M/D}^B/1$ queue to model the process of serving these transactions and subsequently compute the delay.
To do so, we embedded a Markov  chain at departure instants and through method of supplementary variables and PASTA property derived the steady state probability distribution of the system. 
By analyzing real world blob demand in Ethereum blocks, we concluded that less than twice the current demand is required to reach the current target load of EIP-4844.
We further showed that less frequent larger transactions lead to higher delays compared to more frequent smaller ones.
Introducing classes of transactions based on the number of their blobs and incorporating monetary policies like exponential EIP-1559 into our model are left for the future. 

\bibliographystyle{IEEEtran}
\bibliography{IEEEabrv, references}

\begin{thebibliography}{10}
\providecommand{\url}[1]{#1}
\csname url@samestyle\endcsname
\providecommand{\newblock}{\relax}
\providecommand{\bibinfo}[2]{#2}
\providecommand{\BIBentrySTDinterwordspacing}{\spaceskip=0pt\relax}
\providecommand{\BIBentryALTinterwordstretchfactor}{4}
\providecommand{\BIBentryALTinterwordspacing}{\spaceskip=\fontdimen2\font plus
\BIBentryALTinterwordstretchfactor\fontdimen3\font minus
  \fontdimen4\font\relax}
\providecommand{\BIBforeignlanguage}[2]{{%
\expandafter\ifx\csname l@#1\endcsname\relax
\typeout{** WARNING: IEEEtran.bst: No hyphenation pattern has been}%
\typeout{** loaded for the language `#1'. Using the pattern for}%
\typeout{** the default language instead.}%
\else
\language=\csname l@#1\endcsname
\fi
#2}}
\providecommand{\BIBdecl}{\relax}
\BIBdecl

\bibitem{Danksharding}
\BIBentryALTinterwordspacing
``Ethereum roadmap: Danksharding,'' 2024. [Online]. Available:
  \url{https://ethereum.org/en/roadmap/danksharding/}
\BIBentrySTDinterwordspacing

\bibitem{rollup-roadmap}
\BIBentryALTinterwordspacing
V.~Buterin, ``A rollup-centric ethereum roadmap,'' 2020. [Online]. Available:
  \url{https://ethereum-magicians.org/t/a-rollup-centric-ethereum-roadmap/4698}
\BIBentrySTDinterwordspacing

\bibitem{EIP-4844}
\BIBentryALTinterwordspacing
V.~Buterin, D.~Feist, D.~Loerakker, G.~Kadianakis, M.~Garnett, M.~Taiwo, and
  A.~Dietrichs, ``Eip-4844: Shard blob transactions,'' 2022. [Online].
  Available: \url{https://eips.ethereum.org/EIPS/eip-4844}
\BIBentrySTDinterwordspacing

\bibitem{crapisProjection}
\BIBentryALTinterwordspacing
D.~Crapis, ``Eip-4844 fee market analysis: Simulation and backtest,'' 2023.
  [Online]. Available:
  \url{https://github.com/dcrapis/blockchain-dynamic-pricing/blob/main/eip-4844-sim.ipynb}
\BIBentrySTDinterwordspacing

\bibitem{crapis2023eip}
D.~Crapis, E.~W. Felten, and A.~Mamageishvili, ``Eip-4844 economics and rollup
  strategies,'' \emph{arXiv preprint arXiv:2310.01155}, 2023.

\bibitem{park2024impact}
S.~Park, B.~Mun, S.~Lee, W.~Jeong, J.~Lee, H.~Eom, and H.~Jang, ``Impact of
  eip-4844 on ethereum: Consensus security, ethereum usage, rollup transaction
  dynamics, and blob gas fee markets,'' \emph{arXiv preprint arXiv:2405.03183},
  2024.

\bibitem{zhou2020solutions}
Q.~Zhou, H.~Huang, Z.~Zheng, and J.~Bian, ``Solutions to scalability of
  blockchain: A survey,'' \emph{IEEE Access}, 2020.

\bibitem{thibault2022blockchain}
L.~T. Thibault, T.~Sarry, and A.~S. Hafid, ``Blockchain scaling using rollups:
  A comprehensive survey,'' \emph{IEEE Access}, 2022.

\bibitem{EIP-4844-FAQ}
\BIBentryALTinterwordspacing
V.~Buterin, ``Proto-danksharding faq,'' 2022. [Online]. Available:
  \url{https://www.eip4844.com/#faq}
\BIBentrySTDinterwordspacing

\bibitem{l2beat}
\BIBentryALTinterwordspacing
``The state of the layer two ecosystem,'' 2018. [Online]. Available:
  \url{https://l2beat.com/scaling/summary}
\BIBentrySTDinterwordspacing

\bibitem{bhat1964imbedded}
U.~N. Bhat, ``Imbedded markov chain analysis of single server bulk queues,''
  \emph{Journal of the Australian Mathematical Society}, 1964.

\bibitem{cox1955analysis}
D.~R. Cox, ``The analysis of non-markovian stochastic processes by the
  inclusion of supplementary variables,'' in \emph{Mathematical Proceedings of
  the Cambridge Philosophical Society}, 1955.

\bibitem{kleinrock1974queueing}
L.~Kleinrock, \emph{Queueing Systems, Volume I}.\hskip 1em plus 0.5em minus
  0.4em\relax Wiley, 1974, no. v. 1.

\bibitem{github}
\BIBentryALTinterwordspacing
``Eip-4844: Analysis of blob usage in transactions,'' 2024. [Online].
  Available: \url{https://github.com/Pourya-Sol/EIP-4844}
\BIBentrySTDinterwordspacing

\bibitem{EIP-4844-OLD}
\BIBentryALTinterwordspacing
V.~Buterin, ``Blob transactions,'' 2021. [Online]. Available:
  \url{https://notes.ethereum.org/@vbuterin/blob_transactions}
\BIBentrySTDinterwordspacing

\bibitem{BatchCompression}
\BIBentryALTinterwordspacing
``Optimism batch submission,'' 2024. [Online]. Available:
  \url{https://specs.optimism.io/protocol/derivation.html#batch-submission}
\BIBentrySTDinterwordspacing

\bibitem{Ponyo}
\BIBentryALTinterwordspacing
``Shared sequencing network: A middleware blockchain for decentralizing
  rollups,'' 2023. [Online]. Available:
  \url{https://xangle.io/en/research/detail/1217}
\BIBentrySTDinterwordspacing

\end{thebibliography}

\end{document}